\documentclass[journal]{IEEEtran}

\usepackage{amsmath,amssymb}
\usepackage{booktabs}
\usepackage{graphicx}
\usepackage{cite}
\usepackage{url}
\usepackage{array}
\providecommand{\texorpdfstring}[2]{#1}
\graphicspath{{figures/}}
\newcommand{\frb}{\texttt{frb100-40}}
\newcommand{\ULSAp}{ULSA$+$}

\title{\texorpdfstring{\texttt{frb100-40}}{frb100-40} After Two Decades: An Optimality Certificate and a Preregistered Search Study}

\author{Onur~U\u{g}urlu%
\thanks{Onur U\u{g}urlu is with the Department of Computer Engineering, \.{I}zmir Bak{\i}r\c{c}ay University, \.{I}zmir, T{\"u}rkiye. ORCID: 0000-0003-2743-5939.}}

\begin{document}
\maketitle

\begin{center}
\footnotesize\emph{This work has been submitted to the IEEE for possible publication. Copyright may be transferred without notice, after which this version may no longer be accessible.}
\end{center}

\begin{abstract}
For more than 20 years, the Model-RB benchmark \texttt{frb100-40} remained an open challenge; since 2014, its public record had stood at 99 of 100 variables. We give a directly checkable 100-vertex independent set for its 4,000-vertex graph. Together with a verified partition into 100 cliques of size 40, the witness proves that the maximum independent-set size is 100 and the minimum vertex-cover size is 3,900. The stochastic run that found the witness is kept separate from this proof. We evaluated its added pair and triple repair operators in a preregistered campaign comprising 8,668 valid runs. The primary comparison found no detectable acceleration over base ULSA (hazard ratio 0.967, 95\% confidence interval 0.915--1.023; $p=0.248$), and the factorial ablation reached the same conclusion. On a smaller FRB suite, the group-aware CSP pipeline solved 2,500/2,500 runs, compared with 2,391/2,500 for LibMVC-NuMVC. On \texttt{frb100-40}, full ULSA, base ULSA, and NuMVC each produced 0/56 new certificates. With no events, the planned cross-solver hazard ratios remain unidentified. NuMVC ended with cover size 3,902 in 40 runs and 3,903 in 16. Exhaustive enumeration showed that none of the 108 unique recorded conflict-two states had a strictly improving group-aware CSP neighbor within Hamming radius three. The certificate settles the instance. The experiments characterize the search barrier, and the preregistered comparisons show no heuristic advantage.
\end{abstract}


\section{Introduction}
\IEEEPARstart{H}{ard} satisfiable instances provide an unusually direct meeting point between computational-intelligence methods and complexity-motivated benchmark design. Model RB was introduced to generate random constraint-satisfaction problems (CSPs) with growing domains and an analytically located satisfiability transition \cite{xu2000}. Forced-satisfiable instances near that transition can remain difficult for both complete and incomplete search \cite{xu2005}. Their graph encodings, distributed through BHOSLIB \cite{bhoslib}, have consequently become standard tests for maximum clique, maximum independent set (MIS), minimum vertex cover (MinVC), and coloring algorithms.

Among these instances, \frb{} is the largest commonly discussed case, representing 100 variables with domain size 40. Rosin's unweighted stochastic local-search algorithm (ULSA) reached 99 satisfied variables in 2014 \cite{rosin2014}. The benchmark history still listed 99 as the latest public record when we checked it on 2 September 2026 \cite{bhoslib}, and a 2025 paper described the instance as an unsolved 20-year challenge \cite{xu2025}. Accordingly, our deliberately qualified priority statement is that this is the first publicly checkable size-100 assignment known to us.

Once found, a complete assignment is easy to check. It selects one vertex from each of 100 groups and must contain no graph edge. Since each group is a clique, the same file also supplies a matching upper bound. The central result of this paper is consequently a short certificate for a fixed graph hash: a 100-vertex independent set, an independently verified 100-clique partition, and the resulting equalities $\alpha(G)=100$ and $\tau(G)=3900$.

Finding the witness and explaining the search are different questions. The successful run used periodic pair and triple repairs. One rare trajectory provides insufficient evidence for their utility. We therefore replayed the run exactly and tested the operators under a frozen protocol. The resulting 8,668-run campaign includes the smaller-instance comparison, a factorial ablation, and three matched 56-run validation arms on \frb. The preregistered comparisons found no speed advantage for the repairs.

The unsuccessful runs still leave useful structural evidence. Complete enumeration around 108 unique conflict-two assignments shows that none has a strictly improving grouped-CSP neighbor within Hamming radius three. The resulting proposition applies to the archived states. Other states and local-search methods lie outside its scope. It points to a specific barrier: escaping these endpoints requires a neutral or worsening transition, a change involving at least four variables, or a move outside the audited neighborhood.

\section{Background and Related Work}

\subsection{Model RB and the FRB Graph Encoding}

A binary Model-RB instance has $n$ variables, a domain of size $d=n^{\alpha}$, and randomly selected forbidden value pairs \cite{xu2000,xu2005}. The standard graph encoding creates one vertex $(i,a)$ for each variable $i$ and value $a$. All $d$ values of a variable form a clique; a forbidden tuple between variables $i$ and $j$ adds a cross-group edge. A CSP solution is therefore an independent set containing exactly one vertex from every group.

For \frb, $n=100$ and $d=40$, hence the graph contains 4,000 vertices. The supplied DIMACS instance has 572,774 edges. Our parser independently classified 78,000 distinct intra-group edges, exactly $100\binom{40}{2}$, and 494,774 cross-group edges. The intra-group cliques give the immediate upper bound $\alpha(G)\leq100$.

\subsection{Stochastic Local Search for CSP and Vertex Cover}

Stochastic local search balances improving steps with mechanisms that traverse plateaus or escape local minima \cite{hoos2004}. Plateau traversal is also central to maximum-clique search, where dynamic penalties and swaps can sustain exploration after direct improvement stalls \cite{pullan2006}. ULSA is an unweighted, group-aware CSP method designed for random binary instances. It samples a violated constraint and updates one endpoint using vectorized value scores, recency information, and randomized tie breaking \cite{rosin2014}. Its representation always maintains one selected value per CSP variable.

General-graph MinVC methods use a different state space. Edge-weighted local search and configuration checking preceded NuMVC, which combined a two-stage exchange with edge weighting and forgetting \cite{cai2010,cai2011,cai2013}. FastVC later reduced per-step complexity for massive graphs \cite{cai2015}, while MetaVC automated the configuration of a broader collection of MinVC components \cite{luo2019}.

The neighboring MIS literature includes efficient two- and three-vertex exchanges \cite{andrade2012}, evolutionary search coupled with reductions \cite{lamm2017}, and exact branch-and-reduce algorithms for vertex cover and independent set \cite{akiba2016,xiao2017}. Systematic exploration of larger MinVC neighborhoods has shown that exhaustive $k$-swap search can remain practical well beyond the smallest radii \cite{katzmann2017}. This context motivates our repair operators and the radius-three enumeration. It also cautions against attributing a pipeline difference to representation alone because ULSA and NuMVC change both solver and encoding-aware state space.

\section{The Certificate and the Search}

\subsection{A Checkable Proof of Optimality}

Let $V_1,\ldots,V_{100}$ be the consecutive 40-vertex groups in \frb. Direct enumeration verifies that $G[V_i]\cong K_{40}$ for every $i$ and that these groups partition $V$. The argument follows the certifying-algorithm principle: checking the instance, output, and witness is independent of trusting the program that produced them \cite{mcconnell2011}.

\textbf{Proposition 1.} The hashed \frb{} graph has $\alpha(G)=100$ and $\tau(G)=3900$.

\emph{Proof.} Every independent set intersects each clique $V_i$ in at most one vertex, so $\alpha(G)\leq100$. The certificate in Appendix~A contains 100 distinct in-range vertices, one from every $V_i$, and an exhaustive scan of all 572,774 edges finds zero edges with both endpoints selected. Hence $\alpha(G)\geq100$ and therefore $\alpha(G)=100$. For any graph, the complement of an independent set is a vertex cover and $\alpha(G)+\tau(G)=|V|$; thus $\tau(G)=4000-100=3900$. \hfill$\square$

The graph SHA-256 is
\begin{center}
\scriptsize\texttt{1f94968e4506fa91674e5728ff2d0a54}\newline
\texttt{a006e44363302c2c83b188cb2c870066}.
\end{center}
All remaining certificate, log, replay, source, and build digests are recorded in the accompanying artifact manifest.

\subsection{ULSA$+$}

\ULSAp{} leaves ULSA's core update unchanged and adds two periodic strict-improvement operators.

\begin{itemize}
  \item Every $2^9=512$ core updates, a currently violated constraint is sampled. For its two endpoint variables, all $d^2$ joint value assignments are evaluated against the current assignment. The best assignment is applied only if it strictly reduces the total number of violations.
  \item Every $2^{18}=262{,}144$ updates, two variables are obtained from a sampled violation and a third is sampled from violation endpoints (with a bounded fallback). All $d^3$ joint assignments are evaluated with safe lower-bound pruning and, again, only a strict improvement is accepted.
\end{itemize}

The pair and triple variants toggle these operators independently; the full variant enables both. The instrumented executable uses the same observation path in every factorial arm. Before the campaign, the full instrumented build was checked bit-for-bit against the production \ULSAp{} build, and the all-disabled build against Rosin's ULSA for fixed seeds. Sampled logging preserved the solver's random-number consumption. The measured instrumentation overhead in the cluster self-test was approximately 0.12\%.

\subsection{Discovery Run and Deterministic Replay}

The discovery cohort comprised 56 independent \ULSAp{} processes. Task 33, seed 164415777, reached zero violations after 695,602,633,182 core updates, as recorded on the \texttt{bestconflicts 0 iteration} line, and produced the certificate in Appendix~A. Its recorded interval was 26 July 2026 06:19:37 UTC to 28 July 2026 03:45:02 UTC (45.42 h). The other 55 discovery processes ended their budgets without a certificate.

\begin{figure}[t]
  \centering
  \includegraphics[width=\columnwidth]{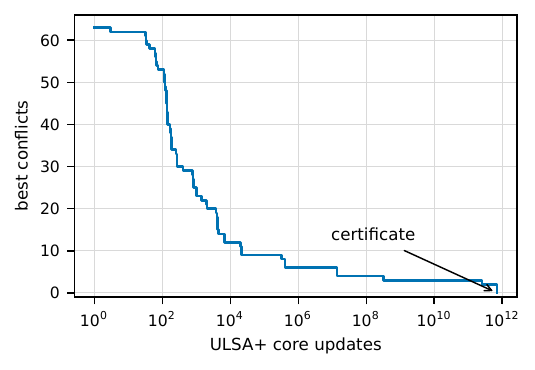}
  \caption{Best violation count in the successful discovery process. The horizontal axis is symmetric-log scaled to retain the early transient and the long conflict-two tail.}
  \label{fig:trajectory}
\end{figure}

The replay used the production source, the same seed, and the full budget. Its solver output was byte-identical to the frozen reference archived with the artifact. The replay demonstrates deterministic same-seed reproducibility. Independent stochastic replication would require new seeds. Fig.~\ref{fig:trajectory} shows the discovery best-so-far trace.

\section{Preregistered Experimental Design}

\subsection{Cohorts, Instances, and Budgets}

The manifest and analysis plan were frozen before the validation results were inspected, and the executable materials were retained to support reproducibility \cite{pineau2021}. The discovery cohort predates this protocol and is treated as exploratory throughout. Table~\ref{tab:cohorts} separates discovery, primary rare-event validation, suite evaluation, and ablation. C, B1, and B2 share 56 seed identifiers as matched initialization blocks. Their subsequent random streams diverge because the repair operators and the NuMVC implementation consume random numbers differently.

\begin{table}[t]
\caption{Preregistered Campaign Structure and Current Status}
\label{tab:cohorts}
\centering
\footnotesize
\begin{tabular}{@{}llllr@{}}
\toprule
Arm & Solver/variant & Purpose & Budget/run & Rows \\
\midrule
Discovery & \ULSAp{} production & certificate search & 71.75 h & 56 \\
C & instrumented full & independent validation & 71.75 h & 56 \\
B1 & instrumented base & independent validation & 71.75 h & 56 \\
B2 & LibMVC-NuMVC & independent validation & 71.75 h & 56 \\
A1 & full/base/NuMVC & suite comparison & 600 s & 7,500 \\
A2 & repair variants & factorial/diagnostic & 600 s & 1,000 \\
\bottomrule
\end{tabular}
\end{table}

A1 contains five instances from each family \texttt{frb30-15} through \texttt{frb59-26}. The first four families use 100 matched seeds per instance and the last four use 25. The development instances \texttt{frb56-25-1} and \texttt{frb59-26-1} were excluded from the primary A1 comparison. A2 uses the remaining four instances in those two families with 25 seeds per factorial cell. Cadence-sensitivity arms are secondary and were excluded from the primary factorial model.

Jobs ran as one-core processes on TRUBA's Orfoz partition. Orfoz nodes contain two Intel Xeon Platinum 8480+ processors (112 cores total) and 256 GB RAM \cite{truba}; the dispatcher allocated 56 one-core tasks per node. ULSA builds used GCC with \texttt{-O3 -mbmi -mavx2 -m64 -funroll-loops}; the LibMVC wrapper used C++14, \texttt{-O3}, and \texttt{-march=native}. Source, binary, manifest, preregistration, instance, and dispatcher hashes prevented mixing results across builds. Scientific time used \texttt{CLOCK\_MONOTONIC}; NuMVC timing began before graph loading. A success flag was created only after a C verifier confirmed both feasibility and the target independent-set size.

\subsection{Preregistered Outcomes and Analysis}

The preregistered primary hypothesis H1 was that full \ULSAp{} has a shorter time-to-target distribution than base ULSA on A1 validation instances. The two-sided test used an instance-stratified Cox model \cite{cox1972} with a cluster-robust sandwich covariance for each instance--seed block. The principal effect size was restricted mean survival time (RMST) at the family cutoff; because every A1 CSP run succeeded, RMST equals the observed mean time.

The A2 $2\times2$ model included pair, triple, and pair-by-triple terms, again using instance-stratified Cox regression with seed-block clustering. Success was to be modeled by clustered logistic regression. All 800 validation observations succeeded, leaving that model unidentifiable. H1, the A2 pair main effect, and the A2 pair-by-triple interaction form a three-test Holm family \cite{holm1979}. Success-rate intervals are exact Clopper--Pearson intervals \cite{clopper1934}. RMST uncertainty uses a fixed-seed, 10,000-repetition two-stage bootstrap: instances are resampled first and matched seed blocks second.

A1's comparison with LibMVC-NuMVC \cite{libmvc} is descriptive. Its unit of comparison is the complete operational pipeline, encompassing representation, algorithm, and implementation. NuMVC medians are computed among successful runs and are displayed together with success rates.

\section{Results}

\subsection{Certificate and Corpus Verification}

An independent parser confirmed 4,000 vertices, 572,774 edge records, 100 complete 40-cliques, 100 unique certificate vertices, and zero selected conflict edges. The archived discovery log and replay output were also checked against the artifact manifest; the replay was byte-identical to its frozen reference.

Across the frozen validation corpus, 8,668 runs were completed: 7,500 A1, 1,000 A2, and 56 in each of C, B1, and B2. All 8,391 emitted success flags were rechecked by the independent C verifier; no unverified success is included. The B2 archive contains exactly the preregistered 56 seeds, and its run records agree byte for byte across the two retained campaign archives. Every B2 process completed normally after its 258,300-s scientific budget; no outer timeout fired.

The successful discovery seed came from an earlier exploratory cohort, so the validation event count excludes it. Its same-seed replay establishes deterministic reproducibility; the stochastic evidence comes from the independently seeded runs.

\subsection{Primary and Factorial Tests}

Table~\ref{tab:cox} summarizes the preregistered Cox-model results. Under the frozen analysis, H1 received no support: the hazard ratio for full \ULSAp{} relative to base ULSA is 0.967 (95\% CI 0.915--1.023; raw $p=0.248$, Holm-adjusted $p=0.743$). With success as the event, a ratio above one would favor faster full-variant completion. The RMST estimates are 2.540 s for full and 2.283 s for base, a difference of 0.257 s (two-stage bootstrap 95\% CI $-0.367$ to 1.134 s) and a ratio of 1.112 (95\% CI 0.836--1.460).

\begin{table}[t]
\caption{Preregistered Cox Models}
\label{tab:cox}
\centering
\footnotesize
\begin{tabular}{@{}lccc@{}}
\toprule
Contrast & Hazard ratio (95\% CI) & Raw $p$ & Holm $p$ \\
\midrule
H1: full vs base & 0.967 (0.915--1.023) & 0.248 & 0.743 \\
A2: pair & 0.966 (0.796--1.173) & 0.728 & 1.000 \\
A2: triple & 0.993 (0.823--1.198) & 0.939 & --- \\
A2: pair$\times$triple & 0.975 (0.762--1.249) & 0.843 & 1.000 \\
\bottomrule
\end{tabular}
\end{table}

Repair effects were likewise unsupported in A2: the pair, triple, and interaction estimates are all close to one with wide intervals. Mean completion times for base, pair, triple, and full were 19.291, 18.219, 16.277, and 16.839 s, respectively. The preregistered stratified model gives no evidence that these marginal differences represent repair effects. All 1,000 A2 runs, including the 200 cadence-sensitivity runs, attributed the final successful transition to ULSA's core update; no pair or triple repair supplied that step.

\subsection{Pipeline Performance on the Smaller FRB Suite}

\begin{figure*}[t]
  \centering
  \includegraphics[width=0.96\textwidth]{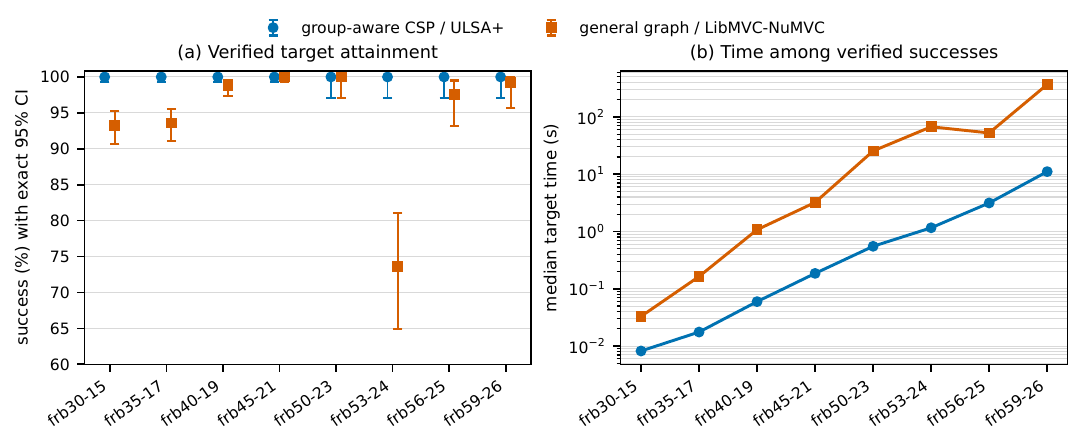}
  \caption{A1 pipeline comparison by FRB family. (a) Verified success proportions with exact 95\% binomial intervals. (b) Median target time among verified successes on a logarithmic scale; the medians exclude censored NuMVC runs. Solver, representation, and implementation vary together; the panels therefore summarize complete-pipeline performance.}
  \label{fig:pipeline}
\end{figure*}

On the A1 suite, full \ULSAp{} produced 2,500/2,500 verified solutions (100\%; 95\% CI 99.85--100\%), while the LibMVC implementation of NuMVC produced 2,391/2,500 (95.64\%; 95\% CI 94.76--96.41\%). Table~\ref{tab:pipeline} and Fig.~\ref{fig:pipeline} break this down by family. The medians use target time among verified successes and exclude censored NuMVC runs. The descriptive family totals include the two development instances excluded from H1. On \texttt{frb59-26}, for example, the full group-aware pipeline solved 125/125 runs with a successful-run median of 11.085 s; LibMVC-NuMVC solved 124/125 with a successful-run median target time of 367.312 s.

\begin{table}[t]
\caption{A1 Pipeline Results by Family}
\label{tab:pipeline}
\centering
\scriptsize
\begin{tabular}{@{}lrr@{\;}rr@{}}
\toprule
& \multicolumn{2}{c}{CSP / full \ULSAp} & \multicolumn{2}{c}{Graph / LibMVC-NuMVC} \\
Family & Success & Med. s & Success & Med. s \\
\midrule
frb30-15 & 500/500 & 0.008 & 466/500 & 0.033 \\
frb35-17 & 500/500 & 0.018 & 468/500 & 0.164 \\
frb40-19 & 500/500 & 0.060 & 494/500 & 1.075 \\
frb45-21 & 500/500 & 0.186 & 500/500 & 3.215 \\
frb50-23 & 125/125 & 0.551 & 125/125 & 25.435 \\
frb53-24 & 125/125 & 1.161 & 92/125 & 67.213 \\
frb56-25 & 125/125 & 3.148 & 122/125 & 52.454 \\
frb59-26 & 125/125 & 11.085 & 124/125 & 367.312 \\
\bottomrule
\end{tabular}
\end{table}

For a practitioner who begins with the original CSP groups, this difference matters: the group-aware pipeline was more reliable and reached its targets sooner on this suite. The comparison covers complete operational pipelines. The general-graph solver searches a broader cover space and also differs in its moves, weighting, initialization, and stopping rule. Isolating representation would require the same search logic under both encodings or a factorial solver-by-representation design.

\subsection{Validation Outcomes on \texttt{frb100-40}}

No validation arm produced a new certificate: C, B1, and B2 each returned 0/56. The exact two-sided 95\% interval is 0--6.38\% for every arm. The absence of events leaves the planned Cox contrasts and hazard ratios unidentified and provides no basis for equivalence or a ranking of the three solvers on \frb.

Those event-free runs still provide useful endpoint information. All 112 C and B1 runs stopped with two violated CSP constraints. NuMVC reported a best cover size of 3,902 in 40 runs and 3,903 in 16, corresponding to objective gaps of $+2$ and $+3$ from the certified optimum. It never reported the target, either within or after the scientific budget.

\begin{figure}[t]
  \centering
  \includegraphics[width=\columnwidth]{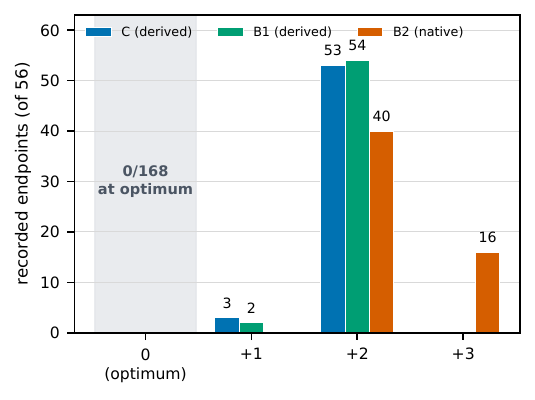}
  \caption{Endpoint objective gaps in the three preregistered validation arms ($n=56$ per arm). No arm produced a verified certificate (0/56 in each; exact two-sided 95\% CI 0--6.38\%), so the planned hazard ratios remain unidentified. C and B1 gaps are reconstructed post hoc from recorded assignments and independently verified; B2 gaps are NuMVC's reported cover values minus 3,900.}
  \label{fig:frb100}
\end{figure}

To compare terminal objective quality in Fig.~\ref{fig:frb100}, we performed a post-hoc conversion of the C and B1 assignments. Each selected one vertex per group and induced exactly two conflict edges. Removing a minimum vertex cover of those two edges yields a valid independent set of size 99 for five records and 98 for the remaining 107; an all-edge rescan verified every derived set. The C and B1 $+1$/$+2$ values are reconstructed from archived assignments. The NuMVC $+2$/$+3$ values come directly from its native cover objective. Their visual agreement at $+2$ is suggestive. Its post-hoc origin and the simultaneous solver changes preclude a confirmatory cross-solver interpretation.

\subsection{Radius-Three Structure of the Recorded Endpoints}

Every C and B1 run recorded a terminal assignment with two violated cross-group constraints. The 112 records contained 108 unique assignments. For an assignment $x\in\{1,\ldots,40\}^{100}$, let $f(x)$ be its number of violated constraints and $d_H$ the number of variables assigned different values.

\textbf{Proposition 2 (computer-assisted, recorded-state scope).} For every unique recorded conflict-two assignment $x$ and every grouped-CSP assignment $y$ with $1\leq d_H(x,y)\leq3$, $f(y)\geq2$.

The audit first recomputed $f(x)=2$ from the graph. Complete pair evaluation covered 887,040,000 joint value assignments over the 112 run records. For triples, any variable triple disjoint from all endpoints of the two current violated edges cannot remove either violation and was eliminated safely. The remaining 2,085,040 variable triples required 133,442,560,000 joint assignments; none improved on two violations. A second implementation independently reproduced both counts and the zero-improvement result. Delta accounting agreed with full conflict recounting on 10,000 deterministic spot checks.

Proposition 2 applies only to the audited conflict-two states and to strict improvements within radius three. Equal-cost, worsening, and larger-radius moves remain possible; indeed, the successful discovery trajectory ultimately escaped this objective level.

The operator logs provide a complementary, exploratory view. Table~\ref{tab:repair} uses pair attempts consistently across conflict levels. In each of three families, acceptance was zero at the lowest conflict level reached in these trajectories. Positive acceptance remained one level above. The estimand is conditional acceptance along the observed trajectories; uniform state-space density would require a different sampling design.

\begin{table}[t]
\caption{Exploratory Pair-Repair Acceptance by Conflict Level}
\label{tab:repair}
\centering
\footnotesize
\begin{tabular}{@{}lrrr@{}}
\toprule
Family & Conflicts & Accepted/attempted & Rate \\
\midrule
frb56-25 & 1 & 0/8,039 & 0\% \\
         & 2 & 9,989/209,415 & 4.770\% \\
frb59-26 & 1 & 0/33,188 & 0\% \\
         & 2 & 30,992/254,928 & 12.157\% \\
frb100-40 & 2 & 0/9,426,240 & 0\% \\
          & 3 & 4,762,255/52,693,991 & 9.038\% \\
\bottomrule
\end{tabular}
\end{table}

\begin{figure}[t]
  \centering
  \includegraphics[width=\columnwidth]{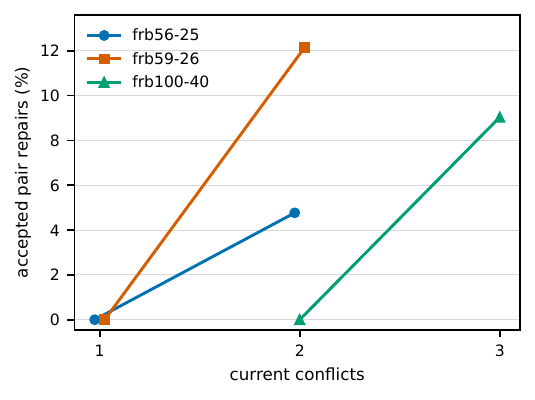}
  \caption{Pair-repair acceptance at each recorded conflict level. Exact counts appear in Table~\ref{tab:repair}. Only \frb{} received an exhaustive radius-three state audit.}
  \label{fig:repair}
\end{figure}

The 112 conflict-two records had mean Hamming distance 95.991 (range 89--100) from the particular optimum certificate in Appendix~A. Two independent uniform assignments over 40 values differ in expectation on 97.5 of 100 variables. Thus, the terminal records are far from this \emph{particular} witness despite having objective value two. Their distance to the nearest optimum remains unknown because other optima may exist.

\section{Discussion}

The certificate is unconditional for the hashed input: anyone can verify its 100 vertices, zero internal edges, and the 100-clique cover. The proof depends only on the witness and clique cover. This separation protects the central result from the low empirical discovery rate.

The preregistered results provide no evidence that pair and triple repairs explain the discovery. They yielded no detectable time-to-target gain over base ULSA, and every final transition in the A2 successes came from the core update. The evidence supports a narrower interpretation: the certificate arose from a rare trajectory generated by a reproducible search system.

The recorded endpoints say more than the aggregate failure count. At each conflict-two endpoint, a strictly improving move must either change at least four CSP variables at once or pass first through an equal- or higher-conflict assignment. That statement is exact for the archived states. Neutral walks, strategic worsening, large-neighborhood search, and nonlocal proposals are therefore natural mechanisms to test next.

On the smaller suite, the ULSA pipeline performed better. The source of that advantage remains unidentified. The A1 comparison changes the solver, state space, and implementation together, leaving the contributions of the CSP representation, ULSA update, vectorization, initialization, and their interactions unresolved. The result informs pipeline selection; a causal encoding claim would require a controlled design. MetaVC appeared in a prospective phase-2 commitment and remained outside the frozen A1 study. Its addition would require a separately labeled experiment.

Several limitations bound these interpretations. The successful seed came from an exploratory 56-run cohort, making its discovery rate unsuitable as a preregistered estimator; the same-seed replay contributes only deterministic evidence. Each validation arm observed 0/56 successes. This all-censored outcome leaves the planned cross-arm hazard ratios unidentified and provides no basis for solver equivalence.

The radius-three result has a similarly finite scope. The audit uses terminal assignments from 56 matched seeds in each of two arms. After accounting for duplicate states and the matched-seed design, the evidence consists of 108 unique archived states with dependence across records. The proposition applies to this finite set of states and the audited neighborhood.

The landscape analysis was designed after the main campaign. Its radius-three enumeration, endpoint-gap reconstruction, and operator-level cross-family summary were independently checked. These analyses remain exploratory. The concentration of both pipelines near objective gap $+2$ is descriptive and is confounded by different state spaces and endpoint definitions. Exhaustive neighborhood enumeration was performed only for the recorded \frb{} CSP states; the smaller-family pattern comes from logged operator attempts.

Execution history introduces one further qualification. C and B1 completed in different calendar blocks; the preregistration had scheduled them for the same week. A required v2.4 dispatcher/preregistration amendment was logged after the campaign began, although solver source and binaries were unchanged. The artifact retains the amended preregistration, final dispatcher, and recorded source and binary hashes. Node and calendar effects remain possible for C versus B1. The stronger A1 and A2 conclusions rely on interleaved matched blocks and instance-stratified analyses.

Finally, reproducibility depends on preserving the complete computational record. The working package contains the certificate, compact results, scripts, and audit reports. A persistent archive must also retain the raw logs, exact source, build environment, manifests, verifier, and hashes. The priority search should be refreshed immediately before submission.

\section{Conclusion}

The 100 vertices reported here settle the hashed \frb{} instance. They contain no internal edge, and the graph's partition into 100 cliques supplies the matching upper bound. Thus $\alpha(G)=100$ and $\tau(G)=3900$ independently of how the witness was found. The successful trajectory is also reproducible: its same-seed replay gives identical solver output.

The search study produced a mixed result. Across 8,668 completed preregistered runs, pair and triple strict-improvement repairs gave no detectable improvement over base ULSA. The group-aware CSP pipeline performed better than LibMVC-NuMVC on the smaller FRB suite, with solver, representation, and implementation varying together. On \frb{} itself, all three 56-run validation arms were event-free, so cross-solver hazard ratios and rankings are unavailable. Complete enumeration establishes a finite structural result: 108 archived conflict-two states are closed to every strictly improving grouped-CSP move within radius three. Methods designed for this benchmark should therefore look beyond the audited neighborhood or accept neutral and worsening steps deliberately.

\section*{Data and Code Availability}

The complete reproducibility artifact contains the hashed graph identifier, the 100-vertex certificate, independent verifiers, discovery and replay logs, the frozen manifest, amended preregistration, final dispatcher, canonical results, analysis scripts, raw campaign logs, and build receipts. It has been deposited at Zenodo under the reserved DOI \url{https://doi.org/10.5281/zenodo.22257064}; the record will be activated upon acceptance.

\section*{Acknowledgment}

The large-scale computational campaign reported in this article was carried out on TRUBA, the national high-performance computing infrastructure operated by T{\"U}B{\.I}TAK ULAKB{\.I}M. The authors thank TRUBA for providing the computational resources. In accordance with the IEEE policy on AI-generated content, the authors disclose that generative AI systems (Anthropic Claude and OpenAI GPT) were used, under author direction, to assist in drafting and revising the manuscript text and in developing and cross-checking the analysis and verification code. All experiments were executed by the authors, and every scientific claim, number, and certificate in this article was verified and approved by the authors, who take full responsibility for its content.

\appendices
\section{Independent-Set Certificate}

The following one-based DIMACS vertices form the verified independent set:

{\scriptsize\ttfamily
25 59 96 142 164 240 248 311 341 392 422 457 483 528 591 613 657 685 757 794 830 847 903 943 992 1015 1067 1105 1151 1174 1227 1264 1314 1357 1381 1432 1470 1499 1557 1586 1618 1663 1710 1729 1786 1819 1868 1909 1952 1968 2038 2076 2104 2130 2174 2234 2247 2293 2333 2371 2408 2446 2496 2536 2566 2612 2668 2696 2733 2788 2832 2863 2907 2956 2963 3036 3080 3112 3121 3185 3227 3280 3311 3333 3382 3429 3463 3485 3545 3563 3603 3680 3688 3733 3764 3818 3873 3888 3958 3987.
}

\bibliographystyle{IEEEtran}
\bibliography{references}

\end{document}